\documentclass[10pt, conference, compsocconf]{IEEEtran}
\usepackage{cite}

\ifCLASSINFOpdf
  \usepackage[pdftex]{graphicx}
  \graphicspath{{..}}
  \DeclareGraphicsExtensions{.pdf,.jpeg,.png}
\else
\fi

\begin{document}
%
\title{Prevalence, Contents and Automatic Detection of KL-SATD}


\author{\IEEEauthorblockN{Leevi Rantala, Mika M{\"a}ntyl{\"a}}
\IEEEauthorblockA{ITEE, M3S\\
University of Oulu\\
Oulu, Finland\\
Email: leevi.rantala@oulu.fi, mika.mantyla@oulu.fi}
\and
\IEEEauthorblockN{David Lo}
\IEEEauthorblockA{Information Systems\\
Singapore Management University\\
Singapore, Singapore\\
Email: davidlo@smu.edu.sg}
}


%


\maketitle

\begin{abstract}
When developers use different keywords such as TODO and FIXME in source code comments to describe self-admitted technical debt (SATD), we refer it as Keyword-Labeled SATD (KL-SATD). We study KL-SATD from 33 software repositories with 13,588 KL-SATD comments. We find that the median percentage of KL-SATD comments among all comments is only 1,52\%. We find that KL-SATD comment contents include words expressing code changes and uncertainty, such as remove, fix, maybe and probably. This makes them different compared to other comments. KL-SATD comment contents are similar to manually labeled SATD comments of prior work. Our machine learning classifier using logistic Lasso regression has good performance in detecting KL-SATD comments (AUC-ROC 0.88). Finally, we demonstrate that using machine learning we can identify comments that are currently missing but which should have a SATD keyword in them. Automating SATD identification of comments that lack SATD keywords can save time and effort by replacing manual identification of comments. Using KL-SATD offers a potential to bootstrap a complete SATD detector. 

\end{abstract}


%
\IEEEpeerreviewmaketitle

\let\svthefootnote\thefootnote

\section{Introduction}
\let\thefootnote\relax\footnote{Copyright © 2020 IEEE.  Personal use of this material is permitted. Permission from IEEE must be obtained for all other uses, in any current or future media, including reprinting/republishing this material for advertising or promotional purposes, creating new collective works, for resale or redistribution to servers or lists, or reuse of any copyrighted component of this work in other works.}
\addtocounter{footnote}{-1}\let\thefootnote\svthefootnote
Technical debt is a term used to depict non-optimal choices made in the software development process. Several types of technical debt has been identified such as code debt, design and architectural debt, environmental debt, knowledge distribution and documentation debt, and testing debt \cite{yan2018automating}. Self-admitted technical debt (SATD) refers to a specific type of code debt, where the developer acknowledges admitting code debt into the system~\cite{potdar2014exploratory}. This admittance can be done in several ways, such as writing a comment into the code or explaining it in a commit message.

One specific way of marking SATD on a code level is to leave a comment with a specific keyword such as TODO. In this paper we focus on four different keywords, which are TODO, FIXME, HACK and XXX. These have all been referred in previous literature as indicators of self-admitted technical debt when present in code comments \cite{de2015contextualized, storey2008todo, maldonado2015detecting}. We call SATD messages labeled with one of these keywords as Keyword-Labeled SATD (KL-SATD). We note that not all SATD comments have these or any keywords. 

In this paper we performed an empirical study on 33 repositories containing more than 500,000 comments to answer the following research questions about SATD comments:
\begin{itemize}
    \item \textbf{RQ1:} \textit{What is the prevalence of KL-SATD comments?} To answer this question, we detect comments for source code and analyze how many of them contain a SATD keyword.
    \item \textbf{RQ2:} \textit{Do the contents of KL-SATD comments differ from other comments?} For this question, we perform a word distribution analysis for all comments, and examine if the comments marked with KL-SATD differ from other comments by their vocabulary.
    \item \textbf{RQ3:} \textit{Can we automatically find comments that have omitted SATD keyword?} Here we use KL-SATD data to train a machine learning classifier but use it to detect SATD comments that have omitted the keyword. This can be beneficial as developers do not necessarily use keywords consistently in their code comments. 
    This SATD detector can be seen as automated labeler that can be trained cheaply from existing SATD comments. Automating this step in SATD tooling is important, as it eliminates the manual labor, which is both time consuming and prone to errors.
\end{itemize}

The paper is structured as follows. Section~\ref{sec:methodology} describes the methodology, starting from the used dataset, continuing with the processing of the comments, and ending with particularities pertaining the research questions
.  Section~\ref{sec:results} shows the results, starting from KL-SATD prevalence, then looking into contents of KL-SATD comments, and ending with automatically finding comments that have omitted SATD keyword. Section~\ref{sec:validity} talks about limitations pertaining to our work, and finally Section~\ref{sec:conclusion} presents the conclusions of the study.

\section{Methodology}\label{sec:methodology}

\begin{figure*}
\centerline{\includegraphics[width=\linewidth]{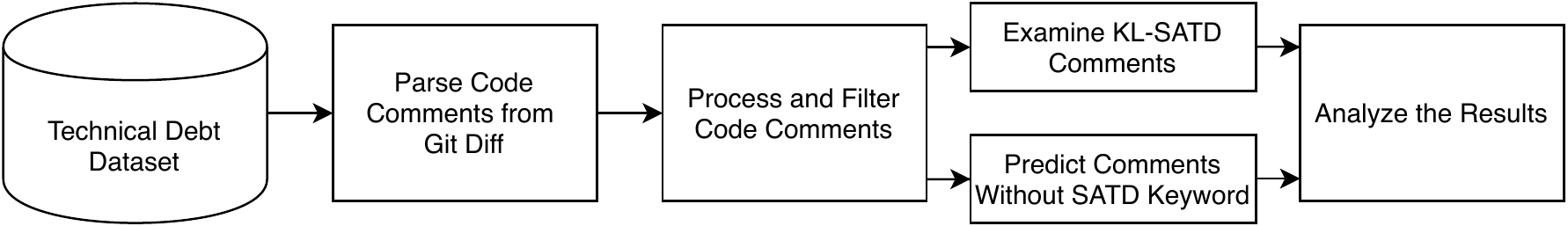}}
\caption{Overview of the methodology}
\label{fig:methodology_overview}
\end{figure*}

The overview of the methodology is shown in Figure~\ref{fig:methodology_overview}.

\paragraph{Data.} We utilize the Technical Debt Dataset created by Lenarduzzi et al.~\cite{lenarduzzi2019technical}. It includes 33 Java projects, which are all over three years old, have more than 500 commits, and include over 100 classes. 
The dataset has a total of over 140 thousand commits in it. We extract both multi- and single-line comments by identifying Java-style patterns with Regex. In total we extracted 862,342 comments.

\paragraph{Preprocessing.} 
The parsed code comments contain lots of noise, and we reduce it processing the comments using following steps: First, we remove comments that are code using NLoN tool ~\cite{mantyla2018natural}. Second, similar to earlier works~\cite{yan2018automating, da2017using}, we deleted all comments that contained license or copyright information, and Javadoc comments containing tags. It has been shown earlier that these comments are not likely to contain SATD~\cite{maldonado2015detecting}. Third, we remove comments that appear in renamed files by excluding all the comments that were deemed as renamed ones according to the Git log in the Technical Debt Dataset. Fouth, all messages were tokenized; next, we transformed all the words to lower case, removed non-alphanumeric characters, and deleted possible email and website addresses. Then we filtered out words that were less than three characters long. Fifth, we filtered out stop words. Sixth, even with stop word removal, some noise was still left to the dataset such as misspelled words, meaningless words, and even some very rare terms like ``abstractauthenticationtoken''.  We solved this by choosing that a word has to appear at least in 5 repositories before it is included in the final vocabulary. This removes even further specific words relating only to couple of projects, and gets rid of many of the misspelled words. Important benefit is that it makes the vocabulary and our results more generalizable. Preprocessing reduces the total number of comments to 507,254.

\paragraph{RQ1} asks about prevalence of KL-SATD comments. To answer this, we calculated on how many commits new comments appeared, and how many of these comments contained SATD keyword. This enables us both to compare projects with each other, and form a general conception of KL-SATD prevalence.

\paragraph{RQ2} looks into how the contents of KL-SATD comments differ from other comments. For this paper, we examined the differences visually by creating comparison word clouds based on word frequencies between KL-SATD comments and other comments. This allows us to see whether they differ in their contents.

\paragraph{RQ3} uses a machine learning classifier to detect SATD comments. We selected logistic regression with lasso penalty from Glmnet package\footnote{https://cran.r-project.org/web/packages/glmnet/index.html} as our machine learning approach. It has been shown to have fast performance, and to work with large and sparse matrices~\cite{friedman2010regularization} which are typical in NLP tasks. It has been applied successfully to different NLP tasks~\cite{genkin2007large}. Logistic regression performs automatically feature selection, and prevents overfitting with penalty term lambda. We perform stratified 10-fold cross-validation and report the results using lambda, which gives the maximum mean for area under the receiver operating characteristics curve (AUC-ROC). We chose AUC-ROC as it performs well with unbalanced datasets~\cite{prati2011survey}. To further alleviate class imbalance, we assign weights to the training data so that KL-SATD comments weigh as much as the non-KL-SATD comments. This equal weighting has improved performance when predicting defects from imbalanced datasets~\cite{herbold2013training}.
We performed term frequency–inverse document frequency (tf-idf) transformation for the comments before entering them to the classifier. This common procedure ensures that common terms are given less value than rarer terms.

\section{Results}\label{sec:results}

\subsection{RQ1. What is the prevalence of KL-SATD comments?}

Total number of comments from all projects after preprocessing was 507,254, from which 13,588 were KL-SATD comments. Descriptive statistics are in Table ~\ref{table:repo-comments-todo-comments}.  The mean of KL-SATD comments per repository was 2.29\%, while median was 1.52\%. The largest percentage of KL-SATD comments were found on commons-bcel project, on which 8.78\% (n=954) from all 10,867 comments present on that project were KL-SATD comments. The least amount in terms of percentage was found on commons-fileupload, which had 2 KL-SATD comments out of 690 comments in total (0.29\%).

\begin{table}
    \centering
    \caption{Summary of all comments and KL-SATD \newline comments per repository}
    \label{table:repo-comments-todo-comments}
 \begin{tabular}{rrrr}
  \hline
            & All       & KL-SATD      & KL-SATD\\ 
            & Comments  & Comments  & Percentage \\
  \hline
  Min.      & 565       & 2         & 0.29 \\ 
  1st Qu.   & 2,849     & 25        & 0.90 \\ 
  Median    & 4,567     & 68        & 1.52 \\ 
  Mean      & 15,371    & 411.8     & 2.29 \\ 
  3rd Qu.   & 10,931    & 324       & 3.05 \\ 
  Max.      & 112,780   & 2,943     & 8.78 \\ 
  \hline
  Total     & 507,254   & 13,588 &
  \\
   \hline
\end{tabular}
\end{table}

We further explored why a repository starts getting a higher KL-SATD comment percentage using a Spearman's rank correlation test. We find a low positive correlation between KL-SATD percentage and the following project variables: number of commits (r=0.18), number of developers (r=0.38), and lines of code (r=0.21).  We looked the project duration, and found a low negative correlation (r=-0.25) between development time in months and KL-SATD percentage. As a conclusion, projects with larger size, more developers and which have shorter development history are weakly correlated with a higher KL-SATD comment ratio. However, as the correlations were quite weak, we can’t define a strong reason why this happens.

Our study shows that KL-SATD comment percentages between projects vary greatly, and the median amount for projects being 1.52\%. Previous study~\cite{wehaibi2016examining} analyzed manually comments of five different projects and found SATD percentage between 2.10 - 3.95\% per project (median 3.41\%). Our study's lower percentage means that simply catching KL-SATD comments with keywords is not enough to catch all the SATD related comments. Manually checking and labeling of the comments that have omitted a SATD keyword is time consuming, which shows the need to detect these comments automatically. 

\subsection{RQ2: Do the contents of KL-SATD comments differ from other comments?}
\label{sec:TODOContents}

\begin{figure}
\centerline{\includegraphics[width=7cm]{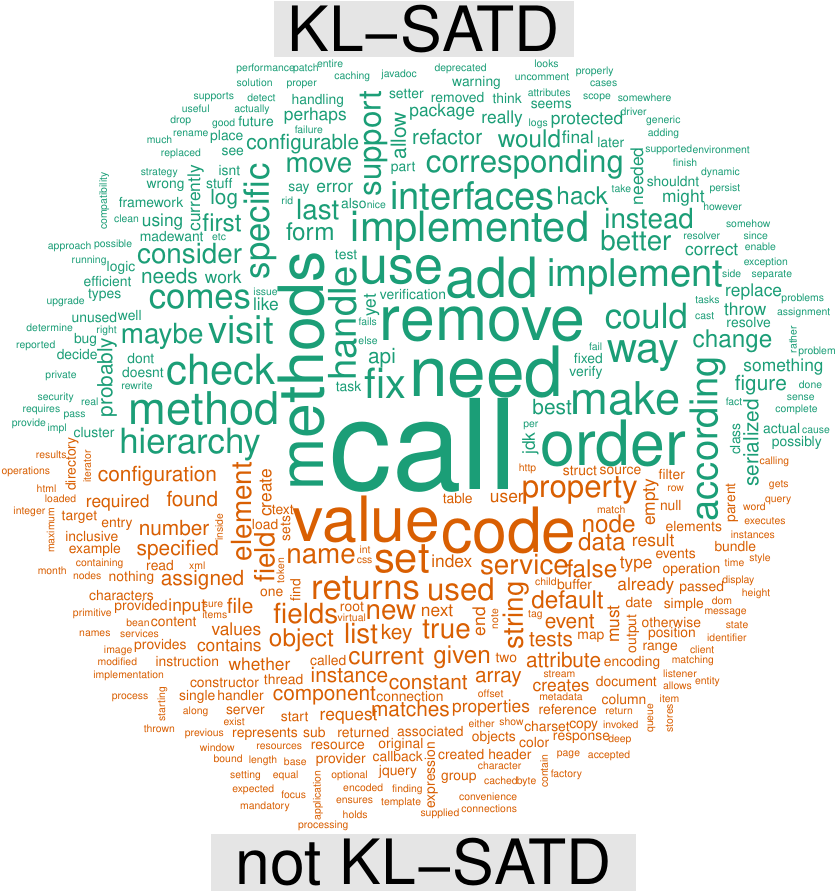}}
\caption{Comparison wordcloud of SATD and non-SATD comments}
\label{fig:wordcloud-comment-contents}
\end{figure}

For analyzing the differences between KL-SATD comments and other comments, we looked at the distribution of words. We created a comparison wordcloud based on the frequencies of words appearing in either KL-SATD comments or non-KL-SATD comments using R-package wordcloud\footnote{https://cran.r-project.org/web/packages/wordcloud/}. Figure~\ref{fig:wordcloud-comment-contents} shows how different words are used in KL-SATD comments (after keyword removal) when compared to other comments.

Words in KL-SATD comments relate to modifying, verifying or adding code functionalities: ``remove'', ``add'', ``need'', ``fix'' and ``check''. KL-SATD comments also include words that show uncertainty such as ``maybe'', ``probably'' and ``consider''. 
Meanwhile, words in comments that are not KL-SATD are more neutral in their tone, and include terms such as ``value'', ``default'', ``code'' and ``element''. Therefore, they seem more like descriptions of code functionality, perhaps related to documenting code.

Looking at the manually labeled SATD comments from previous work~\cite{potdar2014exploratory}, we can see that they share similar vocabulary as our KL-SATD comments. Both contain words such as ``use'',  ``need'', ``remove'', ``implement'', and ``consider''. Even with differences, we can conclude that generic SATD comments and our KL-SATD comments share similarities.

\subsection{RQ3. Can we automatically find comments that have omitted a SATD keyword?}
 
Here we first trained a machine learning classifier using logistic regression to identify comments with KL-SATD. Glmnet logistic regression with stratified 10-fold cross validation gave us an AUC of 0.88 for predicting KL-SATD comments, meaning that our classifier was able to reliably discover which comments had KL-SATD in them with out seeing the keywords.

Examining the classifier's predictor words (Table \ref{tab:5-highest-and-lowest-predictors}) with the most positive and negative coefficients, we see a somewhat different picture than in the word clouds of Section \ref{sec:TODOContents}. This is because we performed tf-idf transformation for the classifier, and did not rely on basic appearance counts. The most positive predictor words include words that are not very visible in the word clouds but ones that you would expect, e.g. refactor, perhaps, consider, and maybe. The most negative predictors include words like ``determined'', ``certain'', ``returns'' and ``sequence''. The full list of predictors can be downloaded from Figshare\footnote{DOI: 10.6084/m9.figshare.11907216}.

\begin{table}
    \centering
    \caption{10 predictor words with the most positive and negative coefficients predicting KL-SATDfrequencies}
    \label{tab:5-highest-and-lowest-predictors}
    \begin{tabular}{r|l|l}
    \hline
            & Positive              &   Negative \\
    \hline
        1. & visit           & certain \\
        2. & consider        & determined \\
        3. & detailed        & concrete \\
        4. & perhaps         & unlike \\
        5. & refactor        & raised \\
        6. & say             & edit \\
        7. & think           & returns \\
        8. & would           & hidden \\
        9. & implement       & casts \\
        10. & fact           & sequence \\
    \hline
    \end{tabular}
\end{table}

Having shown that the classifier is able to detect reliably KL-SATD comments, we then examine the comments that according to the classifier should have KL-SATD in them, but had omitted it. We chose a conservative cutoff point for the predictions, and considered only predictions that were both: a) Not labeled with KL-SATD, and b) Predicted with over 70\% confidence, that they should contain KL-SATD. This gave us 14,690 unique comments that are potentially missing an SATD keyword. To examine the results, we took a random sample of 100 comments, which are availabe from Figshare\footnote{DOI: 10.6084/m9.figshare.12039945}. Two of the authors went over the comments independently, and labeled them with following scale:  0 - Does not contain SATD; 1 - Might contain SATD; 2 - Contains SATD; Empty - Can't tell if SATD or not. The labeling results are presented in Table~\ref{tab:100-random-comments-labeling}.

After labeling, we had 20 different comments with empty label with one or two of the authors. This left us with 80 comments. There were in total 10 comments, which both authors had labeled as 2. These can be considered to be very accurate predictions. In total, there were 23 different comments, which were labeled either as 1 or 2 by both authors, and 24 cases where one reviewer gave label of 0, and then other either 1 or 2. These are more uncertain cases, where the context of the comment (e.g. the code around the comment) might help to make more accurate estimation.
And finally there were 33 cases, where both authors said that the comment does not include SATD.

\begin{table}
    \centering
    \caption{Labeling results of 100 randomly selected comments}
    \label{tab:100-random-comments-labeling}
    \begin{tabular}{l|l|l}
    \hline
    Label       &   Labeler 1   & Labeler 2 \\
    \hline
    0 (Does not contain SATD) &   41          &   60      \\
    1 (Might contain SATD) &   23          &   12      \\
    2 (Contains SATD) &   20          &   18      \\
    Empty (Can't tell if SATD or not)   &   16      & 10       \\
    \hline
    \end{tabular}
\end{table}

In many cases, it was hard to determine whether SATD was present in a particular comment. For example, it is hard to differentiate whether a comment is a description of what the code currently does or what it should do.

Overall, both labelers agreed that on 1/3 of the comments there were no SATD present. This means, that up to 2/3 ($\approx10,000$ comments) of our classifier sample may contain technical debt. However, in many cases more information of the project and source code would be needed to determine the true share of technical debt.

\section{Limitations}\label{sec:validity}
Choice of machine learning algorithm can be seen as a limitation. However, it has been shown that at least in the context of detecting code smells the choice of the algorithm does not give meaningful difference in performance \cite{fontana2016comparing}.

The choice to repositories used in can be seen also as a limitation. We used 33 repositories gathered in a previous study~\cite{lenarduzzi2019technical}, which is a much larger than used in prior works of automated 
SATD detection, e.g. 10 in \cite{ren2019neural} and 7 in \cite{yan2018automating}.

\section{Conclusion and Future Work}\label{sec:conclusion}

In this paper, we studied the prevalence of KL-SATD comments, their contents and automatically finding comments that should include a SATD keyword. All the data and the R-code used to perform the study are available online~\footnote{https://github.com/M3SOulu/KL\-SATD\_SEAA\_2020}. The results show that developers use SATD keywords sparingly (median 1.52\% of comments). We showed that KL-SATD comment contents are different from comments without it, and typically express code changes and uncertainty. KL-SATD comment words are similar to the SATD lexicon shown in prior works. We built a KL-SATD detector that achieved relatively high AUC-ROC of 0.88. Our KL-SATD detector can also find comments where developers have omitted a SATD keyword even though it clearly would need one. Thus, the amount of KL-SATD comments does not accurately reflect the number of comments that should have SATD keyword in them. This points to a problem with the state-of-the-practice 
detectors such as SonarQube, which rely on keyword detection as one part of their technical debt detectors.

We manually labeled only 100 randomly chosen samples, and this does not necessarily reflect the true performance of the classifier. We fully acknowledge this problem in our short paper. To fix shortcomings of our model, we plan to test our classifier with an industry partner (Softagram), and we will also retrain our model with Active Learning methods, which will further enhance its performance.

\bibliographystyle{IEEEtran}
\bibliography{IEEEabrv, bibliography.bib}
%


\end{document}